\documentclass[a4paper,11pt]{JHEP3}
\usepackage{amsmath}
\usepackage{epsf, cite, amssymb}
\usepackage{epsfig}
\title{String Creation, D-branes and Effective Field Theory}
\author{Ling-Yan Hung \\ Department of Applied Mathematics and
  Theoretical Physics, \\ Wilberforce Road, Cambridge CB3 0WA, UK
\\ E-mail: \email{lyh20@damtp.cam.ac.uk} }

\abstract{This paper addresses several unsettled issues associated
  with string creation in
  systems of orthogonal D$p$-D$(8-p)$ branes.
The interaction between the branes can be understood either from the
  closed string or open string picture.
In the closed string picture it has been noted that the DBI action fails to
  capture an extra RR
  exchange between the branes. We demonstrate how this problem
  {\it persists} upon lifting to
  M-theory. These D-brane systems are analysed in the closed string
  picture by using gauge-fixed
  boundary states in a non-standard lightcone gauge, in which RR
  exchange can be analysed precisely.
 The missing piece in the DBI action also manifests itself in the open
  string picture as a mismatch between the Coleman-Weinberg potential
  obtained from the effective field theory and the corresponding open
  string calculation. We show that this difference can be
  reconciled by taking into account the superghosts in the (0+1)
  effective theory of the chiral fermion, that arises from gauge
  fixing the spontaneously 
  broken world-line local supersymmetries.
 }

 \preprint{DAMTP-2008-12}

\begin{document}

\section{Introduction}
It was first observed in \cite{Hanany_Witten} that when a D5 brane aligned along $x^i, i\in\{0,1,2,3,4,5\}$ crosses an NS5 brane aligned along $x^j, j\in \{0,1,2,6,7,8\}$ in the $x^9$ direction, a D3 brane aligned along $x^k, k\in \{0,1,2,9\}$ is created. This is the Hanany-Witten effect\cite{Hanany_Witten}. Examining the NS 3-form flux induced by the NS5 brane on the D5 world-volume leads to the conclusion that a D3 brane has to be created, by virtue of Gauss's theorem, given that the induced flux jumps by one unit as the brane crosses each other.

By a string of T and S dualities, the Hanany-Witten configuration can
be mapped to a system of orthogonal D$p$-D$(8-p)$ branes, either in
type IIA or IIB, oriented so that there is a single spatial direction transverse to the
branes \cite{Bachas}.  The creation of D3 branes
translates into the creation of F1 strings connecting the two branes
as they cross. This phenomenon again directly follows from Gauss's
theorem.  In fact, instead of considering charge conservation in brane
world-volume as in \cite{Hanany_Witten}, we could equally look at the
Gauss's equation of the NS 3-form flux to which fundamental strings
are coupled \cite{Bergman}.  Consider for example a D0 and a D8 brane
in IIA. Gauss's equation for the NS 3-form is given by
\begin{equation}
d \star (e^{-2\phi}dB_2) = 2m \star (dA_1 + mB_2),
\end{equation}
where $B_2$ is the NS 2-form potential, $A_1$ is the RR 1-form
potential under which D0 is charged, and $m$ is the cosmological
constant, dual to $F_{10}$ which couples to D8. Suppose the D8 brane
wraps around an 8-sphere. Integrating both sides of the equation along
the 8-sphere, the l.h.s is identically zero. However, the r.h.s is not
if the D0 brane is enclosed by the D8 due to the contribution of $2m
\star dA_1$. We thus conclude that as the D0 crosses the D8 from
outside the 8-sphere, a string which sources the NS 3-form, has to be
created, giving rise to an extra term $\sim
(2\pi\alpha')^{-1}\delta^{(8)}(x)$ on the r.h.s., which cancels the D0
contribution.  Taking the limit that the radius of the eight-sphere
approaches infinity, we recover the original story of string creation
as the D0 crosses D8. This argument can be readily generalised to
other D$p$-D$(8-p)$ systems, in which case Gauss's equation for the NS
3-form is given by (in type IIA for concreteness)
\begin{equation}
d \star (e^{-2\phi}dB_2) = -\frac{1}{2} F_4\wedge F_4 + F_2\wedge \star F_4 - \star A_1\wedge H_3\wedge F_2,
\end{equation}
where the first two terms on the l.h.s. account for string creation in all the possible D$p$-D$(8-p)$ systems in IIA, including D4-D4 and D2-D6 respectively.  Clearly from T-duality similar equations can be written down for IIB.

This phenomenon, however, raises questions. By open-closed duality,
the interaction between the branes can be analysed either as a closed
string tree level exchange or an open string one-loop potential.
From the closed string picture, a boundary state calculation gives two
pieces, from NSNS and RR exchange respectively, with equal
magnitude. It is given by the product of half of the tension of a string and
separation. However, the RR exchange has a
sign ambiguity. Starting with one convention the sign flips when the
two branes cross each other. Therefore beginning with a vanishing potential
between the branes, a linear potential is created when the branes cross. By
virtue of supersymmetry, which asserts that the force should
vanish, we have to conclude that this nontrivial potential is cancelled by a string
that is created as the branes cross. This is consistent with the above analysis via Gauss's theorem.

Yet this crucial RR exchange is
unexpected since naively the two branes couple to different RR
potentials. The DBI action fails to capture this extra interaction and
thus a probe D$p$ brane in the curved background generated by the
D$(8-p)$ brane experiences a net force due to the gravitons and
dilaton. This is reviewed in section \ref{probebrane}. In the case of
a probe D2 brane in a D6 background the system can be lifted to
M-theory. The D2 brane becomes and M2 brane and the D6 becomes a
Taub-Nut background. The extra duality between the RR
fields then implies a duality 
between the three form potential and the metric. It is then found that
as in the type II limit, the naive membrane action also {\it misses} the 
extra contribution.

A careful analysis in string theory of the unexpected RR exchange
\cite{Divecchia} using the boundary
state formalism suggests that it can be ascribed to an
extra duality
relationship between the time-like modes which are otherwise
decoupled from the string S-matrix calculation in the absence of
boundary sources.  The non-zero overlap between these time-like modes
arises from a cancellation between the zeroes from the trace of Gamma
matrices and divergences from the ghost sector, upon application of a
regularisation procedure. As we will show,  the duality relationship
so obtained is more generic than is described in \cite{Divecchia} and can
be shown to exist between many other components of the potential. We
will also see that the duality relation proposed is neither Lorentz
invariant nor gauge invariant.  These problems will be reviewed in
section \ref{divecchia}.
So the question arises as to how a Lorentz invariant formalism could
have given rise to such a duality relationship. The
main feature of our analysis is the choice of a particular gauge, the
closed string generalisation of the
Arvis gauge \cite{Arvis, Green_gauge}, which is a variant on the light-cone
gauge that avoids the problem of
divergent ghost contributions. We then attack the problem using boundary
methods without the need to confront divergences arising in the ghost
sector. As we shall see, in section \ref{lcgauge}, the same results as
in \cite{Divecchia} are produced. However, this formulation makes it
clear that the duality relations are only an artifact of an explicit
gauge choice. The unexpected RR exchange, is only possible when both
branes are present, where some of the ten-dimensional Lorentz symmetry is broken. Neither
the DBI action nor the supergravity solutions of parallel D-branes,
are affected by these extra duality relations.

The problem can be analysed entirely in the
open string language. In the open string description the interactions
between the branes manifest themselves as a 1-loop Coleman-Weinberg
effective potential after integrating out the open string
ground states that connect the two branes. The open string calculation matches exactly that of
the closed string calculation. The NSNS contribution comes, in the
open string language, from the sum
of two different sectors, namely
\begin{equation}
\frac{1}{2}\textrm{tr}_{\textrm{NS}} q^{L_0-a_{\textrm{NS}}} +
\frac{1}{2}\textrm{tr}_{\textrm{R}} q^{L_0}.
\label{NSNS}
\end{equation}
The RR contribution, comes from
\begin{equation}\label{RR}
\frac{1}{2}\textrm{tr}_{\textrm{R}}
((-1)^{F}q^{L_0}).
\end{equation}

An effective theory calculation however, gives only the
contribution from (\ref{NSNS})
and misses those from (\ref{RR})\cite{Banks}. For consistency and
supersymmetry, the extra
contribution from (\ref{RR}) is put in by hand in the effective theory
and it is known as the \emph{bare term}.
We have seen that in either the closed string or open string description, the
contribution (\ref{RR}) plays a crucial yet mysterious role. It has
also been referred to as a  \emph{half-string} \cite{Bergman,Klebanov}
as an explanation of its stringy origin, since the
term gives rise to a potential  of half that of a string.
The effective theory should provide a microscopic
explanation of the term since it should be possible to map every detail in the
effective theory to the open string calculation. We show that it is a
quantum effect whose
origin could be traced to $\beta\gamma$ ghosts in the
effective field theory description. The discussion is presented in
section \ref{Open}.

\section{D$p$ brane probes in D($8-p$) backgrounds}\label{probebrane}

The general form of a D$p$ brane supergravity solution is
\begin{eqnarray}
ds^2 &=& H^{-1/2}_p dx_{p+1}^2 + H^{1/2}_p dx^2_{9-p}, \nonumber \\
e^{2\phi} &=& H_p^{\frac{3-p}{2}}, \nonumber \\
H &=& 1 + \frac{h_p}{r^{7-p}},
\end{eqnarray}
for some constants $h_p$. Substituting the above metric into the DBI action of a probe Dq brane with eight relatively transverse dimensions and switching off all kinetic terms we readily obtain the potential term given by
\begin{equation}
T_{q} \int d^{q+1} \sigma e^{-\phi} \sqrt{H_p^{\frac{-1+ n-m}{2}}} = T_{q} \int d^{q+1} \sigma  H_p^{\frac{p-3-m+n-1}{4}},
\end{equation}
where there are $m$ shared directions between the D$p$ and D$q$ brane and $p-m+n = 8$ relatively transverse directions.
This gives a potential proportional to
\begin{equation}
V(r) = T_{q} \int d^{q+1} \sigma H_p,
\end{equation}
which is clearly non-vanishing.  This apparent contradiction with
intuition about BPS configurations was observed long ago
\cite{Tseytlin}. Nevertheless, general solutions of probe D$p$ branes
in curved D$(8-p)$ backgrounds have been found\cite{Callan, Craps,
  Camino, Gomis} and are shown to
exhibit the same behaviour as the D2 brane in D6 background, which we
will discuss in further detail in the next subsection.
These solutions are found by first finding a solution corresponding to
the Dq probe wrapping the ($9-p$) sphere transverse to the background
D$p$ brane, with fields depending only on the polar angle $\theta$. In
other words, we solve for the field $r$, the radial coordinate of
the transverse sphere, and the worldvolume electric field $E$ as a
function of $\theta$ only.
Then the brane is decompactified by changing to coordinates
\begin{equation}\label{uncompact}
z = -r \cos{\theta}, \qquad \rho = r\sin{\theta}.
\end{equation}
One aligns the probe brane along $\rho$ instead and obtain $z$ and $E_\rho$ as a function of $\rho$. Starting from the expression for $\frac{r'}{r}$ \cite{Camino} and differentiating (\ref{uncompact}), we finally obtain
\begin{equation}
\frac{dz}{d\rho} = \frac{D_\rho}{\rho^3 H_p},
\end{equation}
where the displacement field $D_\rho = \frac{\partial L}{\partial E_\rho}$.
The probe action can also be simplified to
\begin{equation}
T_q \int dt d\rho H_p \rho^{q-1} \sqrt{1-E_{\rho}^2 + (\frac{dz}{d\rho} )^2}.
\end{equation}
The important lesson from the exercise is that we obtain finally
\begin{equation}
E_\rho = \frac{dz}{d\rho},
\end{equation}
which satisfies the BPS condition required by Kappa symmetry\cite{Bergshoeff},
\begin{equation}
\frac{1}{\sqrt{1+ z'^2 -E^2}} \Gamma_{11}^{p-2/2}[\Gamma_{0\rho\phi_1...} + (\Gamma_{0z}z' + \Gamma_{11}E)\Gamma_{\phi_1...}] \epsilon = \epsilon,
\end{equation}
 as  also observed   in \cite{Gomis} for the case of a probe
 D5 in D3 background.  There are only two independent projections on
 the killing spinors, from the background brane and the first term of
 the above equation and so the system should be quarter BPS.
While the solution is shown to preserve supersymmetry, on closer
inspection it turns out that
there is a net force equal to half of the string tension pulling on
the probe brane. The solution is static since the brane has infinite
extent. This is a general feature of all such D$p$ probe
solutions in  IIA and IIB D($8-p$) supergravity backgrounds\cite{Callan,
  Craps, Camino, Gomis}. We will present an explicit calculation of
 this potential experienced by the D2 probe in the next subsection in the
 context of M-theory.

\subsection{Example in M-theory}
A probe D2 brane in a D6 background can be lifted to
M-theory.
In fact, the first example of an explicit solution of string creation from a
probe brane perspective was given in
\cite{Green_pauli} in the M-theory context. A probe D2 brane is placed
in the curved
background generated by the D6 brane. The system is lifted to M-theory
where the D2 becomes an M2 and the D6, a KK monopole. The curved
metric generated by a KK monopole is equivalent to a Taub-Nut space,
which is a complex manifold and determines a special complex
structure.
The metric of a KK monopole located at the origin can be written, in the Einstein frame, as\cite{Nakatsu}
\begin{equation}
ds^2 = dx_7^2 +  Vdvd\bar{v} + \frac{1}{V}(\frac{dy}{y} - f dv
)\overline{(\frac{dy}{y} - f dv )},
\end{equation}
where $dx_7^2$ denotes the Minkowski metric along the 6+1 dimensional world-volume of the KK monopole and
\begin{eqnarray}
v = \frac{(x^7+ix^8)}{R},  &\qquad& h = \frac{x^9}{R},  \nonumber \\
\sigma = \frac{x^{10}}{R}, \qquad &\qquad& y = e^{-(b+i\sigma)} (-b-+ \sqrt{(b)^2 + |v|^2})^{\frac{1}{2}},\nonumber \\
V= 1 + \frac{1}{2\sqrt{|v|^2 + b^2}}, &\qquad& f = \frac{b+ \sqrt{|v|^2 + b^2}}{2|v|\sqrt{|v|^2 + b^2} }.\nonumber \\
\end{eqnarray}
This background is 1/2 BPS.
Any solution of the membrane action respecting the background complex structure would automatically satisfy a BPS condition and preserves 1/2 of the background supersymmetries.
A supersymmetric M2 brane orthogonal to the KK monopole can thus be
described by a first order holomorphic curve in $v$ and $y$. Higher
order curves would describe multiple M2 branes since they can be
factorised as products of first order curves. There are two such
distinct holomorphic curves, namely\cite{Green_pauli}
\begin{equation}\label{curves}
y = e^{-b}, \qquad y= e^{-b} v.
\end{equation}
Considering their radial profiles it turns out they are mirror images of each other. The IIA limit is taken by sending all length scales including $|x^9|, |v|$ and $b$ to infinity relative to $x^{10}$, while keeping the ratios among them fixed. The parameter $b$ then describes the asymptotic $x^9$ position of the D2 brane. Suppose we consider the first curve in (\ref{curves}) with $b<0$. The D2 brane is located near $x^9 \sim b$ for all $v$. As $b$ is increased the brane is increasingly curved and as $b$ passes through $b=0$, the D2 brane is hooked up at the origin and while it still asymptotes to $x^9 \sim b$ at $v \sim \infty$, near $v\sim 0$, $x^9 \sim 0$. The shape of the curved probe becomes exactly that of a string stretching between the D2 brane at $x^9\sim b$ and the D6 brane at $x^9 = 0$.

To show explicitly that the stretched part is indeed a string, we can calculate the tension of the D2 near $v \sim 0$ for $b>0$.
The energy of the system is simply given by the tension multiplied by the area of the brane. This is given by
\begin{equation}
E = T_2 \int d^2 \zeta \sqrt{\det{G_{ab}}}, \qquad G_{ab} = \partial_a X^\mu \partial_b X^\nu g_{\mu\nu},\end{equation}
where $T_2$ is the brane tension and $G_{ab}$ is the induced metric on the brane.
The complex structure of the background simplifies the problem greatly. The expression of the energy is reduced to
\begin{equation}
E = \frac{T_2}{2} \int dz d\bar{z}   g_{\mu\bar{\nu}} \partial_z X^\mu  \partial_{\bar{z}} X^{\bar{\nu}},
\end{equation}
where $z = \zeta_1 + i\zeta_2$. For concreteness consider the second curve $w = \exp{(-b)}v$\footnote{The second curve is wrapping the 11th direction while the first curve is not. Therefore in the rest of the section we shall restrict our attention to the second curve.}. Since we are aligning the brane asymptotically along $x^7, x^8$ we choose the static gauge
\begin{equation}
z = Rv.
\end{equation}
As a result we have
\begin{equation}
E = T_2 \int \frac{dvd\bar{v}}{2} (V + \frac{f^2}{V}).
\end{equation}
To take the IIA limit and zoom into the region occupied by the extended string, we take
\begin{equation}
R = 1, \qquad x_9, b, |v| \sim +\infty , \qquad x_9 \ll b, \qquad \frac{|v|}{x_9} \ll 1.
\end{equation}
In this limit
\begin{eqnarray}
|v|^2  &\sim& e^{2(x_9 - b)}2x_9, \nonumber \\
|v|^2 + x_9^2 &\sim& 2e^{(-2b)}x_9 + x_9^2 \sim x^2_9, \nonumber \\
V &\sim& 1 + \frac{1}{2x_9} \sim 1, \nonumber \\
f^2 &\sim& \frac{1}{|v|^2}.
\end{eqnarray}
Changing variables from $z$ to $x_9$, we have
\begin{equation}
E = 2\pi T_2 \int e^{-2b(1-x_9/b)} dx_9 (1+2x_9 + 2e^{(2x_9 +b)})\{1+ \frac{1}{2x_9}+ \frac{1}{|v|^2(1+\frac{1}{2x_9})}\} = 2\pi T_2 \int dx_9.
\end{equation}
Given that
\begin{equation}
T_{M2} = \frac{1}{2\pi l_s^2 2\pi g_s},
\end{equation}
and that $g_s = 1$ in our units, the energy in this region of the brane is given finally by
\begin{equation}
\int \frac{1}{2\pi l_s^2} dx_9.
\end{equation}
The result agrees with the fact that a fundamental string connects the D2 to the D6. This argument readily generalises to the case where there are $N$ D6 branes, in which case the modified metric would lead to $N$ strings connecting the branes.
The calculation of the brane energy can be done exactly. Observing the
following relations after using $w= \exp{(-bv)}$ 
\begin{eqnarray}
&&|v|^2 = e^{2(x_9-b)}(2x_9 + e^{2(x_9-b)}) \nonumber \\
&&|r| = \sqrt{x^2_9 + |v|^2}= x_9 + e^{2(x_9-b)}, \qquad f= \frac{|v|e^{-2(x_9-b)}}{2|r|},
\end{eqnarray}
the integral
\begin{equation}
E = 2\pi T_2 \int dx_9 e^{2(x_9-b)}(1+2x_9 +
2e^{2(x_9-b)})\{1+\frac{1}{2|r|} +
\frac{|v|^2e^{-4(x_9-b)}}{2|r|(1+2|r|)}\}. 
\end{equation}
This can be simplified to
\begin{eqnarray}
E&&= 2\pi T_2\bigg\{\int dx_9 \big[\frac{e^{2(x_9-b)}}{2|r|}+e^{2(x_9-b)}
+\frac{|v|^2e^{-2(x_9-b)}}{2|r|}\big]+ \int dx_9 e^{2(x_9-b)}(1+2|r|)\bigg\}
\nonumber \\ 
&&=2\pi T_2\bigg\{\int dx_9
\big[\frac{e^{2(x_9-b)}+x_9+|r|}{2|r|}+e^{2(x_9-b)})\big] + \int dx_9
e^{2(x_9-b)}(1+2|r|)\bigg\}  \nonumber \\
&&=2\pi T_2\bigg\{\int dx_9 (1+2e^{2(x_9-b)}) + \int |v|d|v|\bigg\}.
\end{eqnarray}
The integral is divergent due to the infinite extent of the brane. The
second bracket corresponds to the area of the membrane in a flat
background. Since it is independent of the parameter $b$ we will
ignore it in later expressions. The curve is only defined between
$0<x_9<b$. Therefore,
\begin{equation}
E  = 2\pi T_2\bigg\{\frac{2x_9+e^{2(x_9-b)}}{2}\bigg\}^\infty_0 = 2\pi
  T_2\bigg\{ 
  \frac{|v|^2}{2}[1 + e^{-2(x_9-b)}]\bigg\}^\infty_{0}. 
\end{equation}
To extract the net force on the brane from the infinite energy due to the infinite extent of the brane, we differentiate the expression with respect to $b$ while holding $v$ fixed.
This gives
\begin{equation}
F\equiv \frac{dE}{db}\bigg\vert_v =2\pi T_2 \bigg[\frac{|v|^2}{2} \frac{d
    e^{-2(x_9-b)}}{db}\bigg\vert_v\bigg]^\infty_{0}. 
\end{equation}
Note that
\begin{eqnarray}
\frac{de^{-2(x_9-b)}}{db}\bigg\vert_v &&=
-2e^{-2(x_9-b)}\bigg(\frac{dx_9}{db}\bigg\vert -1\bigg) \nonumber \\ 
&&=-2e^{-2(x_9-b)}\bigg(\frac{2e^{2x_9}+2x_9e^{2b}}{2e^{2x_9}+e^{2b}(1+2x_9)}-1\bigg)
\nonumber \\
&&= \frac{2e^{-2(x_9-b)}e^{2b}}{2e^{2x_9}+e^{2b}(1+2x_9)}. 
\end{eqnarray}

As a result,
\begin{equation}
F = 2\pi T_2
\bigg[\frac{|v|^2}{2}\frac{de^{-2(x_9-b)}}{db}\bigg\vert_v\bigg]^\infty_0 
= 2\pi T_2 \bigg[\frac{e^{2b}(2x_9+e^{2(x_9-b)}
    )}{2e^{2x_9}+e^{2b}(1+2x_9)}\bigg]^\infty_0. 
\end{equation}
In the limit $x_9 \to \infty, x_9 \gg b$,
\begin{equation}
\bigg[\frac{e^{2b}(2x_9+e^{2(x_9-b)}
    )}{2e^{2x_9}+e^{2b}(1+2x_9)}\bigg] \to
    \frac{e^{2b}e^{2(x_9-b)}}{2e^{2x_9}} = \frac{1}{2}.
\end{equation}
On the other hand, in the limit $x_9\to 0, b\to \infty$
\begin{equation}
\bigg[\frac{e^{2b}(2x_9+e^{2(x_9-b)}
    )}{2e^{2x_9}+e^{2b}(1+2x_9)}\bigg] \to e^{-2b}\to 0.
\end{equation}
Putting the pieces together,
\begin{equation}
F= 2\pi T_2 \frac{1}{2}.
\end{equation}
There is thus a {\it net force} acting on the membrane, as already mentioned
at the end of the previous subsection. 
In the next section we shall explore why the potential apparently
does not cancel for a BPS configuration.

\section{Boundary states and extra duality relations}\label{divecchia}
D branes can be represented by boundary states in a closed string theory. Boundary states are BRST invariant states that enforces certain boundary conditions associated with the particular D brane on the string world-sheet ending on it. It is a product of a matter and a ghost part. In the conformal gauge the matter part satisfies the following conditions for a D$p$ brane\cite{Divecchia}
\begin{eqnarray}
\partial_\tau X^l_{\tau = 0} |B_X> = 0, &\qquad&  \partial_\sigma X^t_{\tau = 0} |B_X> = 0, \nonumber \\
\psi^l - i\eta\tilde{\psi^l}|_{\tau = 0} |B_{\psi}, eta> =0, &\qquad&  \psi^t - i\eta\tilde{\psi^t}|_{\tau = 0} |B_{\psi}, eta> =0,
\end{eqnarray}
where $l$ denotes longitudinal directions and $t$ transverse ones. The parameter $\eta = \pm 1$. For an (anti-) brane a unique combination of these states is allowed under GSO projection. A full discussion of the GSO projected states and the ghost states can be found in \cite{Divecchia}.
The interaction energy between two D branes is thus
\begin{equation}
V = <Dp| \Delta |Dq>,
\end{equation}
where $\Delta$ is the closed string propagator (\ref{propagator}) to be discussed in section \ref{lcgauge}.
The NSNS sector gives a finite result\cite{Divecchia}
\begin{equation}
V_{NSNS} = \frac{A}{2\pi}(8\pi^2\alpha')^{-s/2} \int_0^\infty dt (\frac{\pi}{t})^{d/2} e^{-L^2/(2\alpha't)} [(\frac{f_3}{f_1})^{8-\nu}(\frac{f_4}{f_2})^\nu- (\frac{f_4}{f_1})^{8-\nu}(\frac{f_3}{f_2})^\nu],
 \end{equation}
where $s$ is the number of directions shared by the two branes and $d$ is the number of totally Dirichlet directions. The parameter $L$ is the transverse separation of the two branes and $\nu=8$ is the number of relatively transverse directions. The $f_i$'s are the standard combination of Jacobi $\theta$ functions
\begin{eqnarray}
f_1 = q^{\frac{1}{12}}\prod^{\infty}_{n=1}(1-q^{2n}), &\qquad& f_2 = \sqrt{2}q^{\frac{1}{12}} \prod^{\infty}_V(1+q^{2n}), \nonumber \\
f_3 = q^{-\frac{1}{24}}\prod^{\infty}_{n=1}(1+q^{2n-1}), &\qquad& f_4 = q^{-\frac{1}{24}} \prod^{\infty}_{n=1}(1-q^{2n-1}),
\end{eqnarray}
where $q = e^{-t}$.
The RR sector vanishes, however, due to traces of gamma matrices. At
the same time the super-ghost sector also diverges. The product of the
sectors are regularised, by introducing a regulator\cite{Divecchia}
\begin{equation}
x^{F_0 + G_0},
\end{equation}
where $F_0$ and $G_0$ are the fermion number operator and super-ghost
number operator respectively. The regulator is inserted into all
relevant inner products after which the limit $x \to 1$ is taken.  It
then turns out that the order of the zeroes of the trace of the gamma
matrices
$\textrm{tr}(x^{F_0}\Gamma_{11}\Gamma_{i_1}...\Gamma_{i_\nu})$ for
$\nu = 8$  cancels the divergence from the divergence arising from the
super-ghosts. The resultant RR contribution to the interaction energy
is, for $\nu = 8$,
\begin{equation}\label{Ramond_pot}
V_{RR} =  \frac{A}{2\pi}(8\pi^2\alpha')^{-s/2} \int_0^\infty dt (\frac{\pi}{t})^{d/2} e^{-L^2/(2\alpha't)} (\pm 1).
\end{equation}
Adding the contributions, and making use of the abstruse identity
$f_4^8 -f_3^8+f_2^8 = 0$,
\begin{equation}
V = V_{NSNS} + V_{RR} = \frac{V_1}{2\pi\alpha'}L  \qquad \textrm{or} \qquad 0
\end{equation}
depending on the sign chosen in  (\ref{Ramond_pot}), which changes under a parity transformation.
The magnitude of the NSNS and RR exchanges each equals one half string
tension multiplied by brane separation. This is the manifestation of
string creation in perturbative string theory.  On the other hand, the
non-vanishing RR contribution is surprising because naively a D$p$ and
a D$(8-p)$ brane  couples to different RR potentials. To resolve this
apparent contradiction, an ``off-shell'' representation of the RR
potentials was proposed in \cite{Divecchia} to allow for off-shell
momentum exchange between the branes along the totally transverse
directions i.e. only $k_9$ is non-vanishing for the D$p$-D$(8-p)$
configuration.  Combining with the above regularisation procedure it
was found that there is a non-zero overlap between the time-like
components of a $p+1$-form and a $9-p$-form. One of the linear
combinations of the two components are orthogonal to all other states
and the two states could thus be identified.
For example
\begin{equation}
|C_0> = -  |C_{012345678}>,
\end{equation}
which is very much analogous to the more familiar relations
\begin{equation}\label{relation2}
C^{\textrm{transverse}}_p = \star_8 C^{\textrm{transverse}}_{8-p},
\end{equation}
that can also be obtained in this formalism.
It was suggested further that since these extra duality relations
involve only time-like states which decouple from all string
amplitudes, they should not alter the known facts about string
theory. We should note that it is not clear how the duality relations
should manifest themselves in M-theory, which suffers also from a
missing interaction between the M2 brane and KK monopole, as we have
analysed in detail previously.
Moreover, on closer inspection it is clear that the proposed relation
is neither Lorentz invariant nor gauge invariant. Worse still, a D4
brane aligned along $x^{01234}$ should be charged under both
$C_{01234}$ and $C_{05678}$ under this proposal. This calls into
question the validity of the supergravity solutions obtained for all
D$p$ branes in which they are charged only under one component of the
RR potential.

Also, given the off-shell momentum $k_{\mu \ne 0} = 0$, there is no
reason to distinguish the ``transverse components'' from the
``longitudinal components''. In fact, applying the same procedure to all
components $C_i$, $<C_i|C_{012345678}> \ne 0$.  It seems that we might
need to be more cautious about the interpretation of the duality
relations. Recall that the relation (\ref{relation2}) arises naturally
in the light-cone gauge formalisms, where all residual gauge degrees
of freedom are fixed. A light-cone gauge has the advantage of
retaining only the physical degrees of freedom without the need for
ghosts. This suggests the use of light-cone gauge also in attacking
this problem. However, as is already known\cite{Gutperle}, the usual
light-cone gauge is incapable of describing the D$p$-D$(8-p)$
configurations. This is because by default both light-cone directions
become Dirichlet directions whereas the system concerned requires 9
Neumann directions.

\section{Closed string generalisation of Arvis gauge}\label{lcgauge}
\subsection{The bosonic case}\label{bosonic}

The idea of the non-standard light cone gauge makes use of the fact that the left-moving and right-moving modes on the string world-sheet are decoupled from each other for a string theory in a flat background. Therefore we can choose the gauge in each sector independently.

To begin with, the bosonic sector is considered and the gauge choice can be written as

\begin{eqnarray}\label{gauge}
\partial_+ X^+ &=& \alpha' p_+,  \\
\partial_-X^- &=& \alpha' p_-,
\end{eqnarray}

where we have defined
\begin{eqnarray}
\sigma_\pm &=& \tau \pm \sigma , \\
X^{\pm}  &=& \frac{X^0 \pm X^9}{\sqrt{2}}.
\end{eqnarray}
The transverse directions are given by $X^I$ where $  1 \le I \le 8$.
This can be compared with the gauge condition in the usual light-cone gauge
\begin{equation}
\partial_\pm X^+ = \alpha' p_+.
\end{equation}

The gauge condition, together with the equations of motion in the conformal gauge i.e. the plane-wave equation, imply that the light-cone directions admit the following expansions
\begin{eqnarray}
X^+ &=& x_0^+ +  \alpha' p^+ \sigma_+ + \alpha' \hat{p}^+ \sigma_-  + \sqrt{\frac{\alpha'}{2}} \sum_{n\ne 0} \tilde{\alpha}^+_n \exp{(in\sigma_-)},   \nonumber \\
X^- &=& x_0^- + \alpha'\hat{p}^- \sigma_+ +\alpha'p^-\sigma_-  + \sqrt{\frac{\alpha'}{2}} \sum_{n\ne 0} \alpha^-_n \exp{(in\sigma_+)} .  \label{expansion}
\end{eqnarray}
The hat above $p^+ $ and $p^-$ denotes the fact that they are operators to be determined in terms of the transverse modes $\alpha^I_n, \tilde{\alpha}^I_n$.
The gauge conditions have in effect turned off all left-moving $X^+$ oscillations and right-moving $X^-$ oscillations.
With this gauge in place, the Virasoro conditions can be solved
\begin{equation}
T_{++} = \partial_+ X . \partial_+ X = 0= T_{--} = \partial_- X  \partial_- X,
\end{equation}
which gives
\begin{eqnarray}
\alpha^-_n &=& \frac{1}{2\sqrt{2\alpha'}p_+} \sum_m \alpha^I_{m}\alpha^I_{n-m},  \nonumber \\
\tilde{\alpha}^+_n &=& \frac{1}{2\sqrt{2\alpha'}p_-} \sum_m \tilde{\alpha}^I_{m}\tilde{\alpha}^I_{n-m}.
\end{eqnarray}
for $n\ne 0$ and the zero-modes yield
\begin{eqnarray}\label{virasoro}
-2p^+\hat{p}^-  + p^Ip^I  &=&  \frac{1}{2\alpha'}\sum_{m\ne 0} \alpha_{-m}^I \alpha_m^I,  \nonumber \\
-2p^-\hat{p}^+  + p^Ip^I  &=&  \frac{1}{2\alpha'} \sum_{m\ne 0} \tilde{\alpha}_{-m}^I\tilde{ \alpha}_m^I .
\end{eqnarray}

These equations imply that
\begin{eqnarray}
p^+ &=& \hat{p}^+ , \nonumber \\
p^- &=& \hat{p}^-   ,
\end{eqnarray}
which are equivalent to the mass-shell condition and level-matching condition. These conditions are also necessary to make the coefficient of $\sigma$ zero in the Fourier expansion (\ref{expansion}). This ensures that the fields $ X^{\pm}$ are periodic in $\sigma$. This can again be compared to the usual light-cone gauge where the analogous conditions are
\begin{equation}
p^- = \hat{p}^- = \hat{\tilde{p}}^-.
\end{equation}

In this gauge, the generators of Lorentz transformation would be given by
\begin{eqnarray}
J^{+-} &=& x^+\frac{(p^- + \hat{p}^-)}{2} - x^-\frac{(p^+ + \hat{p}^+)}{2},\nonumber \\
J^{-I}  &=&  x^-p^I - x^I \frac{(p^- + \hat{p}^-)}{2}  + \frac{\alpha'}{2} \sum_{n\ne 0} [\alpha^-_{-n}\alpha^I_n -\alpha^I_{-n}\alpha^-_n],\nonumber \\
J^{+I}  &=&  x^+p^I - x^I \frac{(p^+ + \hat{p}^+)}{2}  + \frac{\alpha'}{2} \sum_{n\ne 0} [\tilde{\alpha}^+_{-n}\tilde{\alpha}^I_n - \tilde{\alpha}^I_{-n}\tilde{\alpha}^+_n] ,
\end{eqnarray}
Since the anomalies are cancelled separately among the left-movers and the right-movers, these generators give the correct Lorentz algebra.

\subsubsection{Spectrum}\label{bosonic_spectrum}
The spectrum in this gauge yields the same number of degrees of freedom as in the usual light-cone gauge. Consider the massless closed-string states. They are given by
\begin{equation}
\alpha_{-1}^\mu\tilde{\alpha}^\nu_{-1} |0,k> .
\end{equation}
The symmetric traceless part of which gives the gravitational perturbations $g_{\mu\nu}$ while the anti-symmetric part gives the excitations of the 2-form potential $B_{\mu\nu}$.
Our choice of gauge then gives
\begin{equation}
K_{+I} = K_{I-} = K_{+-} = 0,
\end{equation}
where $K_{\mu\nu} = g_{\mu\nu}+B_{\mu\nu} + \eta_{\mu\nu}\Phi$ and
$\Phi = \textrm{K}/10$ is the dilaton.
Since the conformal gauge on the string world-sheet implies the Lorentz gauge on the potentials, we can solve for the non-physical components by making use of the following
\begin{eqnarray}
-p_-K_{+I} - p_+K_{-I} + p_JK_{JI} &=&0, \nonumber \\
-K_{I+} p_- - K_{I-}p_+ + K_{IJ}p_J &=&0.
\end{eqnarray}

Solving these equations gives
\begin{eqnarray}
B_{+I} &=& - \frac{p_JK_{IJ}}{2p_-}, \nonumber \\
B_{-I} &=&   \frac{p_JK_{JI}}{2p_+}, \nonumber \\
g_{+I} &=&  \frac{p_JK_{IJ}}{2p_-}, \nonumber \\
g_{-I} &=&  \frac{p_JK_{JI}}{2p_+}, \nonumber \\
g_{+-} + B_{+-}+ \delta_{+-}\Phi &=& -\frac{p_Ip_Jg_{IJ}}{2p_+p_-}.
\end{eqnarray}

Substituting these expressions into the action and keeping only the physical d.o.f's it turns out that the non-local terms are cancelled between the Einstein-Hilbert action and the kinetic terms for the 2-form potential. The resultant action is simply
\begin{equation}
S = \frac{p^2}{2} (g_{IJ}g^{IJ} + B_{IJ}B^{IJ} + \Phi^2).
\end{equation}
However, it seems that in this gauge it is impossible to switch off all $B$ and consider pure gravity since $B_{+i}, B_{-i}$ and $B_{+-}$ are also expressed in terms of the transverse components of the metric perturbations.  Now consider the field-strengths $H = dB$ when all $B_{IJ}$ are switched off.  Using the gauge conditions
\begin{eqnarray}
H_{+-I} &=& p_J \frac{-2p_+p_- \delta_{IK}+p_Ip_K}{2p_+p_-} g_{JK},  \nonumber \\
H_{+IJ} &=& \frac{-p_K}{2p_-}(p_J g_{Ik} - p_Ig_{JK}).
\end{eqnarray}
and similarly for $H_{-IJ}$. These equations suggest that in general circumstances there would inevitably be non-trivial $H$.  In systems containing branes with 8 relatively transverse directions a non-trivial $H$ is always induced, as we have seen in section 1. Perhaps this gauge is particularly suited for describing such systems.

\subsection{Superstring generalisation}
\subsubsection{RNS formalism}
So far we have been dealing with bosonic string theory. However this
choice of gauge can readily be extended to superstring theory.
Suppose we work within the RNS formalism and introduce 2d world-sheet
Majorana fermions $\psi^\mu$.
From the equations of motion
\begin{equation}
\psi^\mu = \binom{\psi^\mu_-(\sigma_-)}{\psi^\mu_+(\sigma_+)}.
\end{equation}

{\bf \underline{The NSNS sector}}\\

Anti-periodic boundary conditions imply half-integer modes in the Fourier expansion of the fields.
To be consistent with the gauge choice (\ref{gauge}) we need to set
$\psi^+_+ = \psi^-_- = 0$ in the NS sector, which can be neatly written as
\begin{equation}
\rho ^0 \psi^0 + \rho^1\psi^1 = 0,
\end{equation}
where $\rho^0, \rho^1$ are 2d gamma matrices in a basis as given in \cite{Schwarz}.
The super-conformal constraints $J_{\pm} = \psi .  \partial_{\pm} X =
0$ can then be solved as usual for the left and right movers
separately. \\

{\bf \underline{The RR sector}}\\

In the RR sector, periodic boundary conditions allow integer modes in the Fourier expansion of the fields. The zero modes of the $\psi^\mu$ generate the ten dimensional Clifford algebra.  The ground state of the left movers and right movers are respectively degenerate, each corresponding to a 32 component ten dimensional Majorana spinor.  The GSO projection can be imposed as usual accordingly for type IIA and type IIB string theories. Then we impose further the gauge projection consistent with (\ref{gauge}), namely
\begin{eqnarray}
\Gamma^+ |\Psi_{\textrm{L}} > &=& 0, \nonumber \\
\Gamma^- |\tilde{\Psi}_{\textrm{R}} > &=& 0,
\end{eqnarray}
for the left-moving and right-moving ground states respectively. The super-conformal constraint, which corresponds to the Dirac equation, can then be solved.

\subsubsection{GS formalism}
In order to discuss spacetime supersymmetry, it is more convenient to work with the GS formalism.  To obtain $N=2$ $d=10$ supersymmetry, we introduce the spacetime spinor coordinates $\theta^{a}$, which are the 16-component ten dimensional chiral Majorana spinors, where $a\in {1,2}$. Their chiralities are chosen accordingly in type IIA and IIB superstring theories.

The $\kappa$-symmetry transformations of the spinor coordinates are given by
\begin{equation}
\delta\theta^{a} = 2i \Gamma. \Pi _\alpha \kappa^{a\alpha},
\end{equation}
where
\begin{equation}
\Pi^\mu_\alpha = \partial_\alpha  X^\mu - i\bar{\theta}^{a} \Gamma^\mu \partial_\alpha \theta^a,
\end{equation}
and $\kappa^{a\alpha}$ are 10 dimensional chiral spinors satisfying
\begin{eqnarray}
\kappa^{1\alpha} &=& P_-^{\alpha \beta} \kappa^1_\beta, \nonumber \\
\kappa^{2\alpha} &=& P_+^{\alpha \beta} \kappa^2_\beta.
\end{eqnarray}
where $P_\pm^{\alpha \beta} = 1/(2\sqrt{h}) (h^{\alpha \beta} \pm \epsilon ^{\alpha \beta})$. This implies that $a=1$ describes right-movers while $a=2$ describes left-movers.

Imposing the gauge condition (\ref{gauge}) leads to,
\begin{equation}
\Pi^{+}_+ =  p^+, \qquad \Pi^-_- =  p^- .
\end{equation}
Studying the equations of motion as listed in \cite{GSW} suggests that we can consistently set
\begin{equation}\label{projection}
\Gamma^- \theta^1 =0, \qquad \Gamma^+ \theta^2 =0.
\end{equation}
This choice would keep the gauge choice (\ref{gauge}) unchanged under $\kappa$-transformation of $X^\mu$.
Notice that the same projections were obtained for the ground states in RNS formalism. This should also be compared with the projection adopted in the usual light-cone gauge
\begin{equation}
\Gamma^+ \theta^1 =   \Gamma^+ \theta^2 = 0.
\end{equation}
This will have an impact on the spectrum in the RR sector in this gauge.

The combined supersymmetry and $\kappa$-transformation that preserves the gauge choice is given by
\begin{eqnarray}\label{susytrans}
\delta X^i &=& i\bar{\eta}\Gamma^+\Gamma^i \theta^1 (\frac{p^-}{4}) + i\bar{\eta}\Gamma^-\Gamma^i \theta^2 (\frac{p^+}{4}), \nonumber \\
\delta \theta^1 &=& 2i\Gamma^i\partial_- X^i \eta, \nonumber \\
\delta \theta^2 &=& 2i\Gamma^i\partial_+ X^i \eta.
\end{eqnarray}

The resultant gauge equations of motion again reduce to  $\partial_+\partial _- X^i = \partial_+ \theta^1= \partial_- \theta^2 = 0$.

With the gauge projection condition we can re-write the $\theta^a$'s as eight -dimensional chiral spinors, whose chiralities depends on their chiralities in 10 dimensions. Following conventions in \cite{GSW}, we define
\begin{equation}
S^m  = \binom{\tilde{S}_-^m}{S^m_+} =\binom{\sqrt{2p^-} \theta^{1m}}{ \sqrt{2p^+} \theta^{2m}},
\end{equation}
where $m \in \{1,2,3,4,5,6,7,8\}$ are 8-dimensional chiral spinor indices. The effective action that gives the equations of motion is then
\begin{equation}
\int d^2   \sigma  \frac{-1}{2\pi} (\partial_\alpha X^i \partial^\alpha X^i) + \frac{i}{\pi} (S^{m}\rho^\alpha\partial_\alpha S^m).
\end{equation}
The bosonic fields admit the same Fourier expansion as in the $RNS$ formalism.
The $S^m$ have become world-sheet spinor fields which satisfy periodic boundary conditions.  They admit integer modes in Fourier expansion
\begin{equation}\
S^m = \frac{1}{\sqrt{2}}\binom{\sum_{n} \tilde{S^m_n} \exp{(in\sigma_-)}}{\sum_{n} S^m_n \exp{(in\sigma_+)}}.
\end{equation}
Since $S^m$ satisfy the canonical anti-commutation relations, we have
\begin{eqnarray}
\{S^m_h,S^n_k\} &=& \delta^{mn}\delta_{h+k}, \nonumber \\
\{\tilde{S}^m_h,\tilde{S}^n_k\} &=& \delta^{mn}\delta_{h+k}.
\end{eqnarray}

The super-charges that generate spacetime supersymmetry transformation (\ref{susytrans}) can now be readily obtained.
For concreteness we consider type IIA superstring, in which case $\theta^1$ and $\theta^2$ have the same chirality. The supercharges are
\begin{eqnarray}
Q^m = \sqrt{2p^+}S_0^m, &\qquad & Q^{\dot{m}} = \frac{1}{\sqrt{p^+}} \gamma^i_{\dot{m}m} \sum_n S^m_{-n}\alpha^i_n, \nonumber \\
\tilde{Q}^m = \sqrt{2p^-}\tilde{S}_0^m, &\qquad & \tilde{Q}^{\dot{m}} = \frac{1}{\sqrt{p-}} \gamma^i_{\dot{m}m} \sum_n \tilde{S}^m_{-n}\alpha^i_n.
\end{eqnarray}
The algebra of the supercharges is
\begin{eqnarray}
\{Q^m, Q^n\} = 2p^+\delta^{mn}, &\qquad&  \{Q^{\dot{m}}, Q^{\dot{n}}\} = 2\hat{p}^-\delta^{mn}, \qquad \{Q^m, Q^{\dot{n}}\} = \sqrt{2}\gamma^i_{m\dot{n}}p^i, \nonumber \\
\{\tilde{Q}^m, \tilde{Q}^n\} = 2p^-\delta^{mn}, &\qquad&  \{\tilde{Q}^{\dot{m}}, \tilde{Q}^{\dot{n}}\} = 2\hat{p}^+\delta^{mn}, \qquad \{\tilde{Q}^m, \tilde{Q}^{\dot{n}}\} = \sqrt{2}\gamma^i_{m\dot{n}}p^i,
\end{eqnarray}
where
\begin{eqnarray}
-2p^+\hat{p}^-  + p^Ip^I  &=&  \frac{1}{2} \sum_{m\ne 0} \alpha_{-m}^I \alpha_m^I   + nS^m_{-n}S^m_{n},\nonumber \\
-2p^-\hat{p}^+  + p^Ip^I  &=&  \frac{1}{2} \sum_{m\ne 0} \tilde{\alpha}_{-m}^I\tilde{ \alpha}_m^I + n\tilde{S}^m_{-n}\tilde{S}^m_{n}.
\end{eqnarray}

These are again the mass-shell condition and level-matching condition as in (\ref{virasoro}).

\subsubsection{Spectrum}

Since the zero modes of $S^{m}$ and $\tilde{S}^{\dot{m}}$ satisfy the algebra
\begin{equation}
\{S^m_0, S^n_0\} = \delta^{mn}, \qquad   \{S^{\dot{m}}_0, S^{\dot{n}}_0\} = \delta^{\dot{m}\dot{n}},
\end{equation}
the representation space for the left-movers and right-movers is each given by $8_{\textrm{v}} + 8_{\textrm{c}}$ by triality.  The direct product of the massless vectors again give 64 massless states $g+B$ and the dilaton $\Phi$ in the gauge as discussed in section (\ref{bosonic_spectrum}). The direct product of the massless spinors also gives 64 massless states. However, they decompose into 8 dimensional even forms, as opposed to odd forms obtained in the usual light-cone gauge.

\begin{equation}\label{RR_form}
A_{I_1...I_{2n}} = |\dot{m}> (\gamma^{I_1}...\gamma^{I_{2n}})_{\dot{m}\dot{n}} |\dot{\tilde{n}}>.
\end{equation}
This is a result of our opposite projections on the left-moving and right-moving ten dimensional spinors.
In the usual light-cone gauge the 1-form and 3-form are interpreted as the transverse physical components of the two-form and four-form field strengths in ten dimensions respectively. It is not clear what should be the correct interpretation of these states in (\ref{RR_form}). However, as we shall see in the next section,  an 8d p-form couples to the boundary state corresponding to a Euclidean D$p$ brane (of $p+1$ dimensions with at least 1d embedded along $X^9$).  Therefore a $p$-form $A_{I_1...I_p}$ should probably be interpreted as the $C_{9I_1...I_p}$ component of the $p+1$-form potential.

These forms would satisfy a duality relation as a result of 8d Poincare duality. Namely,
\begin{equation}
A_{+I_1...I_{2n}} = \epsilon_8^{I_1...I_{2n}J_{2n+1}...J_{8}}A_{-J_{2n+1}...J_{8}}.
\end{equation}
This duality condition has been proposed in \cite{Divecchia}. A non-zero overlap between the 2 states concerned is obtained where the infinities from the ghosts are shown to cancel the zero from the trace of gamma matrices by employing a regularisation procedure. In our gauge however this relationship is obtained naturally without any divergence appearing.  Yet it is important to note that the duality relationship following from our gauge is not gauge invariant. It is really our choice of gauge that leads to this particular relation, much in the same spirit of the usual light-cone gauge where a $p$ form is related to an $8-p$ form.
More interestingly, via further Poincare dualities in 10-dimensions, the $C_+$ component is related to the $F_{10}$ form, which is dual to the cosmological constant. It seems that this gauge could be related to massive supergravity.

On the other hand, our interpretation of these states as components of the potentials would lead to an apparent redistribution of degrees of freedom, where the transverse components of the 1-form potential are ``transferred'' to the higher forms.  It is not clear whether this is related to the Stuckelberg symmetry in massive IIA supergravity.

\subsection{Boundary States}
We are now in a position to construct boundary states that represent D-Branes.  The special gauge choice requires automatically that $X^0$ satisfies Dirichlet boundary condition while $X^9$ satisfies Neumann boundary condition. As a result the D-brane is again a Euclidean instanton as in \cite{Gutperle} with at least 1d embedded along $X^9$.

The conditions satisfied by a boundary state representing a D$p$-brane transverse to $X^t$ and along $X^l$ are
\begin{eqnarray}\label{bc}
X^t |B, \eta> = 0, &\qquad& \partial_\tau X^l |B,\eta> =0,  \\
S^m + i\eta M_{mn}\tilde{S}^n |B, \eta> =  0, &\qquad& \tilde{S}^{\dot{m}} + i\eta M_{\dot{m}\dot{n}}\tilde{S}^{\dot{n}} |B, \eta> =0,  \label{superreflect}
\end{eqnarray}
where $\eta = \pm 1$ for brane and anti-brane respectively. The zero-modes of (\ref{bc}) give the combination of supercharges that are preserved by the D$p$-brane while the other orthogonal combinations are the broken supersymmetries. The matrices $M_{mn}, M_{\dot{m}\dot{n}}$ can be solved for as in \cite{Gutperle}. They are given by
\begin{equation}
M_{mn} = (\gamma^1\gamma^2 ...\gamma^p)_{mn}, \qquad   M_{\dot{m}\dot{n}} = (\gamma^1\gamma^2 ...\gamma^p)_{\dot{m}\dot{n}}.
\end{equation}

The conditions (\ref{bc}) can be decomposed into Fourier modes to give
\begin{eqnarray}\label{reflect}
\alpha_n + T. \tilde{\alpha}_{-n} |B, \eta>  = 0, &\qquad&\nonumber\\
S^m_q + i\eta M_{mn}\tilde{S}^n_{-q} |B, \eta> =  0, &\qquad& S^{\dot{m}}_q + i\eta M_{mn}\tilde{S}^{\dot{n}}_{-q} |B, \eta> =  0
\end{eqnarray}
where
\begin{equation}
T = \left(\begin{array}{cc}
-I_{p+1} & 0 \\
0            &  I_{p}
\end{array}
\right).
\end{equation}
The reflections of the transverse modes in (\ref{reflect}) automatically lead to
\begin{equation}
\alpha^-_n + \tilde{\alpha}^+_{-n} |B, \eta> =  0.
\end{equation}
This implies
\begin{equation}
\alpha^0_n + \tilde{\alpha}^0_{-n}   |B, \eta> =  0, \qquad \alpha^9_n - \tilde{\alpha}^9_{-n}   |B, \eta> =  0.
\end{equation}
The D$p$-brane thus lie along $X^9$ transverse to $X^0$ as claimed.

The boundary state satisfying the required conditions is then given by \cite{Gutperle}
\begin{equation}\label{boundarystate}
|B> = \exp(\sum_{q>0}(\frac{1}{q}T_{IJ}\alpha^I_{-q}\tilde{\alpha}^J_{-q})-i M_{mn}S^m_{-q}S^n_{-q}) |B_0>,
\end{equation}
where
\begin{equation}
|B_{0(I_1...I_p,9)}> = [ C(T_{IJ}|I> |J> + iM_{\dot{m}\dot{n}}|\dot{m}> |\dot{n}>)]_{p=0}.
\end{equation}

This implies that the boundary state couples to the RR-forms as
\begin{equation}
<C_{+I_1...I_p}| B_{0(I_1,...I_p,9)}>  = \textrm{tr}(I_8).
\end{equation}
The duality relation discussed above also ensures that a D$p$ brane and a D$(8-p)$ brane couples to the same $p+1$ form, which is exactly the result obtained in \cite{Divecchia}.

\subsection{Tree-level interaction between boundary states}

In order to discuss interactions between two branes, a closed string propagator is needed.  We have observed in section (\ref{bosonic}) that the coefficient of the linear $\sigma$ term in the expansion of the $X^0, X^9$ fields only vanish when both the mass shell condition and level matching condition are satisfied. This raises concern over consistency of introducing off mass shell internal states. On the other hand, for $\partial_\sigma X^0 = \partial_\tau X^9 = 0$ at the boundaries we need only
\begin{equation}
p^+ + \hat{p}^- - \hat{p}^+ - p^- = 0,
\end{equation}
which is satisfied for $p^9 = 0$ and that the level-matching condition is satisfied.  This is indeed the case for the brane configurations under consideration. For the moment we impose the level-matching condition and assume that it is possible to consider off-mass-shell states in a consistent manner.

A closed string propagator is given by \cite{GSW}
\begin{equation}\label{propagator}
\Delta = \int \frac{d^2z}{|z|^2}  z^{L_0}\bar{z}^{\tilde{L_0}}.
\end{equation}
The interaction between boundary states is then given by
\begin{equation}\label{interact}
<B_{p+1}| \Delta | B_{q+1}>,
\end{equation}
where $| B_{q+1}>$ is as given in (\ref{boundarystate}).

If the intermediate state is off-mass-shell, we can write
\begin{equation}
L_0 = p^+ (\hat{p}^- - p^-) = - p^+p^- + \frac{p^Ip^I}{2} + N^{\perp} ,
\end{equation}
and similarly for the right-movers.
Evaluating the zero- modes part of (\ref{interact}), it is given as in usual covariant quantisation. The bosonic sector, for example, would give
\begin{equation}
<p=0, I_L| <p=0, J_ R| T_{IJ} \delta^{d_\perp}(\hat{x}_i) |z|^{\alpha'p^2/2} \delta^{d_{\perp}'(\hat{x}_i - y_i)} |p=0, H_L> |p=0, K_ R>T'_{HK}.
\end{equation}
Expressing the $\delta$ functions as integrals the above expression can be simplified to
\begin{equation}
W \int \frac{d^{d_{\perp Dp + Dq}}Q}{(2\pi)^{d_{\perp Dp + Dq}}} |z|^{\alpha'Q^2/2}\exp{(iQ.y)} \textrm{tr}{(TT')},
\end{equation}
where $W$ is some normalisation constant proportional to the volumes of the branes and we have used
\begin{equation}
<p, I|q, J> = 2\pi \delta(p-q) \delta_{IJ}.
\end{equation}
The non-zero modes involve only the transverse components and can be evaluated as usual. In the special case where the two boundary states have 8 relatively transverse dimensions, the non-zero modes give
\begin{equation}
\frac{f_2^8(|z|)}{f_2^8(|z|)},
\end{equation}
The bosonic zero-modes are proportional to $\textrm{tr}(-I_8) = -8 $ and the fermionic zero-modes are proportional to
 \begin{equation}\label{fermionic}
 \textrm{tr}(\gamma^1 ...\gamma^8) = \pm 8.
 \end{equation}
 The sign depends on the projection of the ten dimensional chiral
 spinor in equation (\ref{projection}).  Upon a parity change along
 $X^0$ we exchange $\Gamma^+$ and $\Gamma^-$ and this would switch the
 sign of equation (\ref{fermionic}). This is in support of the
 well-known fact that when the two branes cross each other a string
 has to be created.
This also implies that D8 brane's coupling to the $C_9$ potential
 changes sign under parity. This is in agreement with the result
 \cite{Townsend} that D8 brane behaves like a domain wall in massive
 IIA SUGRA across which the cosmological constant jumps. It is perhaps
 beyond perturbative string theory
 to see how the cosmological constant jumps in steps as we cross more
 than two D8 branes. This is because each closed string world-sheet
 can at most connect to two boundaries.

\section{The no force condition in the low energy effective field
 theory}\label{Open}
So far we have been analysing the system in the closed string channel
in which the branes interact via tree level closed string
exchange. The same system, however, can be described by open-strings,
which we will discuss in the next subsection. This will lead us to
consider the effective field theory in section \ref{effective}.

\subsection{Open-string description of the D$p$-D$(8-p)$ system}
The closed string tree level interaction potential between the branes
is equivalent to the open-string one-loop Coleman-Weinberg effective
potential by a modular transformation. These open-strings concerned
are stretched between the two sets of branes. According to the
boundary conditions satisfied by the world-sheet supersymmetry
current, they are divided into two sectors, the NS-sector if
anti-periodic and the
R-sector if periodic. The Coleman-Weinberg potential is given by $\log \det
(\partial^2 + M^2) = \textrm{tr} \log (\partial^2 + M^2)$. The trace is turned
into an integral over all momenta, and also runs over
all states in the R and NS sector after GSO
projection. The integral
over momenta can be turned into an integral over the Schwinger
parameter $t$. The potential is most easily evaluated in the
light-cone gauge (or the Arvis gauge for the system concerned) with
all the unphysical states removed. The calculation is reviewed in
appendix \ref{arvis} where the potential is given by
(\ref{coleman_weinberg}). The
projector $(1+(-1)^{F_0}\Gamma_{\textrm{R,NS}})$ inserted in the trace
implements the GSO projection explicitly. The traces without the Gamma
matrices in the R and NS sector i.e. $\frac{1}{2}\textrm{tr}_{\textrm{NS}} q^{L_0-a_{\textrm{NS}}}$ and
$\frac{1}{2}\textrm{tr}_{\textrm{R}} q^{L_0}$, adds up and contribute to a term
\begin{equation}\label{pot1}
\frac{HK}{2\times 2\pi \alpha'} |L|
\end{equation}
in the effective potential, where $H,K$ are the number of branes on
each stack and $L$ the separation.
As long as the branes
are not parallel to each other, there exist directions parallel to one
set of the branes and transverse to the other. In which case, the NS
sector fermions along those directions carry zero modes, so that
$\textrm{tr}_{\textrm{NS}}\Gamma_{\textrm{NS}}=0$ along those directions.
In all other supersymmetric
D-branes configurations, the branes share more than two common
longitudinal or transverse directions. Therefore, for strings stretched
between the two sets of branes, after removing the light-cone
directions there still exist world-sheet fermions having the same
boundary conditions on both end-points and carry zero modes. As a
result $\textrm{tr}_{\textrm{R}} \Gamma_{\textrm{R}} = 0.$ This
however is not true for D$p$-D$(8-p)$ systems in which there are
exactly one common longitudinal direction and one common transverse
direction. Modes along those directions can be removed after gauge
fixing. Along all other directions the world-sheet fermions do not
carry zero modes and $\Gamma_{\textrm{R}}=1$ in the light-cone
gauge, thus giving a non-vanishing contribution to the
Coleman-Weinberg potential. In the conformal gauge however, this term is
resulted from the cancellation between the zero arising from the
world-sheet fermionic zero modes and the divergence from ghost zero
modes\cite{Bergman}, which is explained below\footnote{The situation
  is similar to that in the closed string RR exchange calculation.}.
In fact it contributes to a potential
\begin{equation}\label{open_pot}
\frac{HK}{2\times 2\pi \alpha'}L.
\end{equation}
This extra piece that is absent in all
other supersymmetric
brane configurations can be identified with the unexpected RR exchange
in equation (\ref{Ramond_pot})
under modular transformation in the closed string channel discussed in
previous sections.

The physical interpretation of this term is not clear. As we will
discuss further in the next subsection, it does not have a
microscopic origin in the low energy effective field theory where only
the physical spectrum predicted in string theory is taken into
account. So far this term has been added in by hand to make the theory
consistent and supersymmetric. But a direct correspondence between
string theory and the effective field theory must imply that the term
could be understood entirely in the effective field theory framework.
We therefore would like
to identify the precise string states contributing to this
term in order to give it a microscopic interpretation in the effective
field theory.

The partition sum of all the sectors in (\ref{full_sum}) is a constant
independent of the parameter
$t$. This implies that it receives contribution only from the
world-sheet ground states. To find out the precise ground states,
including possible
contributions from non-physical states in the R sector, we should work
again in the conformal gauge. We must combine
the contributions of the
(R,NS) and (R,R) sector i.e. the two parts that constitute the GSO
projection, rather than viewing them
independently. In other words, we are interested in the world-sheet
fermions and
$\beta\gamma$ zero modes contribution to the partition sum
\begin{equation}\label{zeromode}
F_{\textrm{R}} = F_{\textrm{(R,NS)}} \pm F_{\textrm{(R,R)}},
\end{equation}
where
\begin{eqnarray}
&&F_{(\textrm{R,NS})} =
  \frac{1}{2}\textrm{tr}_{\textrm{R}}((-x)^{G_{\beta\gamma}}q^{2N}),
\qquad F_{(\textrm{R,R})} =
  \frac{1}{2}\textrm{tr}_{\textrm{R}}((-y)^F(x)^{G_{ 
  \beta\gamma}}q^{2N}), \nonumber \\
 && q = \exp{(-\pi t)}.
\end{eqnarray}
A priori the term $F_{\textrm{R,R}}$ is ill-defined due to divergent
sums coming from the infinite number of ghost contributions. The
quantities $x,y$ regulate these divergences\cite{Bergman, Divecchia}
and in the end we must set 
$x,y = 1$. We will soon argue that consistency also requires $x=y$.
The notation
used here and ghost quantisation in the
conformal gauge are explained
in appendix \ref{conformal}. The $\pm$ sign in
(\ref{zeromode}) reflects the ambiguity in 
defining the GSO projection. Without loss of generality we consider
the case where there is exactly one brane on each stack.
Using
the results in appendix \ref{conformal} and keeping track explicitly of
the contribution of the two world-sheet fermion ground states
$|\downarrow>,|\uparrow>$ in the R sector, the zero mode
contribution of the first term in (\ref{zeromode})
is
\begin{equation}
F^{(0)}_{\textrm{(R,NS)}} = \frac{<\uparrow|\uparrow> +
  <\downarrow|\downarrow>}{2}\frac{1}{1+x}. 
\end{equation}
Since $ <\uparrow|\uparrow> = <\downarrow|\downarrow>=1$, $F_{\textrm{(R,NS)}} =1/2$ in the limit mentioned above.
The second term in (\ref{zeromode}) is more subtle because it involves
the divergence of the ghost zero modes.
It is given by
\begin{equation}
F^{(0)}_{\textrm{(R,R)}} = \frac{<\downarrow|\downarrow> -
  y<\uparrow|\uparrow> }{2}\frac{1}{1-x}. 
\end{equation}

This expression is highly constrained by requiring consistency with
Gauss' law. This requires that the value of the potential should jump
by one unit of string tension as one brane crosses the other, as in
the case in \cite{Hanany_Witten}. In other
words the potential must jump by one unit when the sign of the GSO
projection in (\ref{zeromode}) is flipped. This implies that
$F_{\textrm{(R,R)}}$ should take the value $1/2$, which, in turn requires
us to choose 
\begin{equation}
x=y \to 1.
\end{equation}

Since
supersymmetry requires vanishing of the potential between the branes,
the open string calculation again implies the creation of a string as
the branes cross, to cancel the effect of the jump in potential. 
Closer inspection of $F_{\textrm{R}}$ reveals the precise states that
are circulating in the loop. 
Consider choosing the upper sign in equation (\ref{zeromode}) as a
concrete example. We will substitute $y=1$ directly since it will not
affect our following discussion. We will, however, be keeping the sum
as a polynomial in $x^k$ where the index $k$ is equal to the ghost
number. The zero mode contribution to the partition sum
$F^{(0)}_{\textrm{R}}$ is then given by 

\begin{eqnarray}
F^{(0)}_{\textrm{R}}=&& \frac{1}{2}\sum_{n=0}^\infty (-x)^n (<\uparrow|\uparrow> + <\downarrow|\downarrow>)
\pm x^n (-<\uparrow|\uparrow> + <\downarrow|\downarrow>)  \nonumber \\
=&& -\frac{1}{2}\sum_{n=0}^\infty
x^n \Big(<\downarrow|\downarrow>(1 \pm (-1)^n ) - <\uparrow|\uparrow>(1 \mp (-1)^n)  \Big) \nonumber \\
=&&-\sum_{n=0}^\infty
\Big(x^{2n}<\downarrow|\downarrow> - <\uparrow|\uparrow>x^{2n+1}  \Big).
\end{eqnarray}

Out of these states only the $n=0$ term $<\downarrow|\downarrow>$ came from the physical state $|\downarrow>$, which is
annihilated by $\beta_0$. The rest have non-zero ghost excitations of
 $2n$ or $2n+1, n>0$. The lowest physical state is thus a single
fermion, as expected. The rest are two infinite towers of ghost states
of \emph{opposite} chiralities\footnote{Chirality is understood from the
  perspective of the
  T-dual theory where the branes intersect over 1+1
  dimensions.}. Changing the choice of GSO projection
 flips the chirality of the lowest physical fermion and swaps
the two ghost towers.

\subsection{Local supersymmetry in quantum mechanics}\label{effective}

As we have seen, the closed string
tree level calculation of the interaction between the branes can be
recast entirely in the open-string language.  The potential energy
should correspond to the Coleman-Weinberg potential in effective field theory.  However, while the open and
closed string calculations agree, they clearly do not agree with a
naive calculation in field theory, with field content consisting of the
physical ground states of the open strings.
Without loss of generality, consider the low energy effective field theory of a
D0 and $K$ D8 branes. This system is consistent only in the context of
type I' theory in the presence of two orientifold planes. But we could
always take the limit in which 8 D8 branes and their images lie on each of the
orientifold planes and the planes are widely separated. Then we can consider
taking $K$ D8 branes from one of the orientifolds and stay
sufficiently far from either of the orientifold planes.
The effective action of the system is then that of  $(0,8)$ quantum
mechanics given as in \cite{Banks,
  Bachas2}. In the presence of a single D0 and $K$ D8 branes, there are
$K$ ``chiral fermions'' inert under the unbroken
supersymmetries\footnote{Again these are related to chiral fermions in
  $1+1$ dimensions by a T-duality.},
whose action is
\begin{equation}\label{fermionic_action}
\mathcal{L}_{\sigma} =  \sum_r^K [-i \sigma^{\dagger}_r\dot{\sigma}_r -
  \sigma^{\dagger}_r(Y + A_0)\dot{\sigma}_r ],
\end{equation}
where $A_0$ is the difference between the gauge fields on the D0 and
the D8 branes. The fermions are in the fundamental representation of
$U(K)$. The field $Y= Y^{\textrm{D0}}-Y^{\textrm{D8}}$ is the
separation between the branes. Integrating
these fermions out gives\cite{Banks}
\begin{equation}\label{pot}
\frac{K}{2} |A_0 + Y|,
\end{equation}
which is a result of the fact that for fermions, only the retarded
Green's function could be defined in a $0+1$ dimensional field
theory. In other words, this is the energy of $K$
fermionic harmonic oscillators each occupying the filled state, where $(Y + A_0)$ is treated as the
natural frequency of the oscillators. Comparing with the open string
calculation, (\ref{pot}) corresponds to the contribution from
$\frac{1}{2}\textrm{tr}_{\textrm{NS}} q^{L_0-a_{\textrm{NS}}}$ and
$\frac{1}{2}\textrm{tr}_{\textrm{R}} q^{L_0}$ as in equation (\ref{pot1}).  There is an extra
piece in the string calculation, coming from
\begin{equation}\label{extra}
\frac{1}{2}\textrm{tr}_{\textrm{R}}(-1)^{F_0}\Gamma_{11} q^{L_0},
\end{equation}
which gives a term $ \frac{K}{2}(A_0 + Y)$, as discussed before equation (\ref{open_pot}).
From the point of view of effective field theory, the origin of this
term is not immediately clear just by inspecting the physical
spectrum.  It is known just that $ K/2(A_0 + Y)$ is needed to render the theory
consistent and supersymmetric. It is called the \emph{bare} term and
is usually added in by hand\footnote{We thank D. Tong for pointing out
  that the appearance of these terms can be understood as
  a result of ``integrating the chiral fermions in'', starting with a
  manifestly supersymmetric theory with no potential terms. }.
By inspecting the open string calculation in detail in the previous
subsection, we concluded that this term arose from the zero modes of
two sets of $\beta\gamma$ superghosts. These ghosts must
 also emerge naturally in the effective field theory,
since each state in string theory corresponds to some
space-time particle, and the assertion extends to the ghost sector. For
example, the level one massless $bc$ ghosts states $b_{-1}|0>$ and
$c_{-1}|0>$ in open bosonic string theory indeed correspond to the
Fadeev-Popov ghosts  needed for removing the non-unitary components of
the massless gauge vector. The two towers of ghost states that we have
found must also correspond to the infinitely degenerate ground
states of two sets of $\beta\gamma$ ghost fields with masses of
opposite signs in the effective field theory.
The $\beta\gamma$
ghosts are needed in local
super-reparametrisation symmetry on the world-line of the R-sector
fermionic 
particle. Note that this is a target space symmetry i.e. a symmetry of
the $0+1$ dimensional effective field theory and should not be
confused with the string world-sheet supersymmetry. However, given
that the fermionic particle is actually the lowest mode of a
macroscopic stretched string, its 
effective action can be thought of as following from
dimensional reduction of a two dimensional string world-sheet. This
explains the presence of local supersymmetry on its worldline.


To demonstrate the presence of these ghosts, we would like to
construct the action of the Ramond fermions so that it is explicitly
invariant under local super-reparametrisation. Upon gauge fixing the
matter part of the action should reduce to equation
(\ref{fermionic_action}). There will be new terms in the action for
the $\beta\gamma$ ghosts of the form $\beta G\gamma$, where $G$ is a
differential operator that generates the transformation of the
gravitini. Without loss of generality we consider
$K=1$. The coupling to the gauge fields look like a mass term and in
general it would take the form $e f \bar{\sigma}\sigma$, where $e$ is
the world-line metric and $f$ will eventually be identified with $Y +
A_0$. Since in general $\delta_{\epsilon}e \sim i\epsilon \chi$ under
supersymmetry transformation, where $\chi$ is the gravitino, the
variation of the mass term can be cancelled by a term proportional to
$\chi ^2 \bar{\sigma}\sigma $, if
\begin{equation}
\delta_{\epsilon}\chi \sim \dot{\epsilon} - i f \epsilon.
\end{equation} We shall now derive this transformation rule for the
gravitini and justify the claim using superspace method. It is
important to note that the second term in this transformation is
crucial since
it gives rise to the coupling between $f$ and the $\beta \gamma$
ghosts upon gauge fixing the gravitini, as is suggested by the string
calculation. Such transformations for the gravitini are familiar in
gauged supergravity theories.  The gauge fixed local supersymmetry is
generally spontaneously broken in these theories since further global
supersymmetry transformations generally take the gravitini away from
the gauge condition. Also the presence of a term  proportional to
$\chi ^2$ in the action can only arise when there is more than one
supersymmetry such that $\chi$ is complex.  In fact we need exactly two local supersymmetries with
a gauged $U(1)$ R-symmetry.

To set the scene we would first review the superspace formalism in super quantum mechanics with two supercharges\cite{Brink}.
The superspace  has three coordinates $(\tau, \theta, \bar{\theta})$. The super-vierbein fields $E^{A}_{M}$ satisfy
\begin{equation}
E^{M}_{B}E^{A}_{M} = \delta^{A}_{B}.
\end{equation}
The indices $M = (\tau, \theta, \bar{\theta})$ are curved space indices and $A = (t, a, \bar{a})$ are tangent space indices. The super-vierbeins transform as
\begin{equation}
\delta_{\zeta}E^{A}_{M}= \zeta^N\partial_{N}E^{A}_{M} + \partial_{M}\zeta^{N}E^{A}_{N},
\end{equation}
and a general scalar super-field $\Phi$ transforms as
\begin{equation}\label{scalarsuper}
\delta_{\zeta}\Phi= \zeta^N\partial_{N}\Phi.
\end{equation}
Also, under tangent space rotations,
\begin{equation}
\delta E^{a}_{M} = - E^{t}_{M}\phi^{a} + E^{b}_{M}\epsilon^a_{b} T,
\end{equation}
where $a$ here corresponds to any of the two tangent space fermionic indices, and $\phi^a$ and $T$ are fermionic and bosonic parameters respectively, of the $U(1)$ transformations. The other components of the super-vierbein do not transform, nor does a scalar super-field.  To limit the  invariance to the usual supergravity form of super quantum mechanics, the super-vierbein should be partly gauge fixed so that they satisfy
\begin{equation}
E^{t}_M = \Lambda \bar{E}^{t}_{M}, \qquad E^{a}_{M} = \Lambda^{1/2}\bar{E}^a_{M},
\end{equation}
where $\bar{E}^A_M$ is the flat vierbein given by
\begin{equation}
\bar{E}^A_M = \left(\begin{array}{ccc}
1 & 0 & 0\\
-\frac{i}{2} \bar{\theta} & 1& 0 \\
-\frac{i}{2} \theta & 0&1
\end{array}\right),
\end{equation}
and $\Lambda$ is
\begin{equation}
\Lambda = e + \frac{i}{2} (\theta\bar{\chi} + \bar{\theta} \chi) + \bar{\theta}\theta \frac{f}{2}.
\end{equation}
The residual symmetries that leave the gauge choice invariant are
\begin{eqnarray}
\zeta^{\tau} &=& a + \frac{i}{2}(\theta\bar{\beta} + \bar{\theta} \beta), \nonumber \\
\zeta^{\theta} &=&  \beta + \frac{1}{2}\theta \dot{a} -i\theta t + i \theta \bar{\theta} \frac{\dot{\beta}}{2}, \nonumber \\
\zeta^{\bar{\theta}} &=&  \bar{\beta} + \frac{1}{2}\bar{\theta} \dot{a} +i\theta t - i \theta \bar{\theta} \frac{\dot{\bar{\beta}}}{2}, \nonumber \\
T &=& t+ \frac{1}{2}(\theta\dot{\bar{\beta}}- \bar{\theta}\dot{\beta}) + \frac{1}{4}\theta\bar{\theta}\ddot{a}, \nonumber  \\
\phi^\theta &=& \Lambda^{-1/2}\zeta^{\theta}, \qquad \phi^{\bar{\theta}} = \Lambda^{-1/2}\zeta^{\bar{\theta}},
\end{eqnarray}
where $a, \beta, \bar{\beta}, t$ are arbitrary functions of $\tau$ and correspond to general coordinate transformations in $\tau$, local supersymmetry and local $U(1)$ respectively. The field $\Lambda$ transforms as
\begin{equation}
\delta_{\zeta}\Lambda=\zeta^M\partial_{M} \Lambda + \partial_{\tau} \tilde{\zeta}^{\tau}\Lambda,
\end{equation}
where
\begin{equation}
\tilde{\zeta}^{A} = \zeta^{M} \bar{E}^A_{M}.
\end{equation}
As we see, $\Lambda$ transforms with an inhomogeneous term in addition to a homogeneous transformation as a usual scalar super-field.

Then we would introduce the fermionic chiral super-field $\Psi$ satisfying
\begin{equation}\label{chiral_superfield}
D_{\bar{\theta}} \Psi = 0,
\end{equation}
where $D_{A} = \bar{E}^{M}_A \partial_{M}$ and in particular
\begin{equation}
D_{\bar{\theta}} = \partial_{\bar{\theta}} + \frac{i}{2} \theta\partial_{\tau}.
\end{equation}
In component form,
\begin{equation}
\Psi = \psi + i \theta F + i\theta\bar{\theta} \eta.
\end{equation}
Solving the condition (\ref{chiral_superfield}) gives $\eta = \dot{\psi}/2$. Eventually we would gauge fix $\Lambda$ such that $(e, \chi,f) = (1,0, Y+A_0)$.  The correct action for $\Psi$ that reduces to the form (\ref{fermionic_action}) is
\begin{equation}\label{fermionic_action2}
S = \int d\tau d^2\theta \Lambda \Psi^{\dagger} \Psi = \int d\tau \left[-ie\psi^{\dagger}\dot{\psi} + eF^{\dagger}F +\frac{1}{2}(\chi^{\dagger}\psi F-\chi\psi^{\dagger}F) + \frac{f\psi^{\dagger} \psi}{2} \right].
\end{equation}
Clearly $F$ is an auxiliary field and after removing it via the equations of motion the action simplifies to
\begin{equation}
S = \int d\tau \left( -ie\psi^{\dagger}\dot{\psi} + \frac{f\psi^{\dagger}\psi}{2} + \frac{\bar{\chi}\chi\psi^{\dagger}\psi}{4e}\right).
\end{equation}
The action contains the $\chi^2$ term as anticipated at the beginning of the section. However, equation (\ref{fermionic_action2}) also makes it clear that $\Psi$ does not transform simply as in (\ref{scalarsuper}) since $\Lambda$ transforms non-trivially. To find the correct transformation rule for $\Psi$, consider $(\delta_{\zeta} \Lambda)\Psi^{\dagger}\Psi$. This can be written as
\begin{eqnarray}
&&(\delta_{\zeta} \Lambda)\Psi^{\dagger}\Psi = (\tilde{\zeta}^{A}D_{A}\Lambda + \partial_{\tau} \tilde{\zeta}^{\tau}\Lambda )\Psi^{\dagger}\Psi \nonumber \\
&&= \partial_{\tau}(\tilde{\zeta}^{\tau}\Lambda \Psi^{\dagger}\Psi) -
  D_{a}(\tilde{\zeta}^{a}\Lambda \Psi^{\dagger}\Psi) + \Lambda\left[
    \dot{\tilde{\zeta}}^{\tau}
    -\dot{\tilde{\zeta}}^{\tau}
    +(D_{a}\tilde{\zeta}^a)- \tilde{\zeta}^{A}D_{A} \right](\Psi^{\dagger}\Psi).
\end{eqnarray}
From these expressions, we conclude that the action (\ref{fermionic_action2}) is invariant up to total derivative terms if $\Psi$ transforms as
\begin{equation}
\delta_{\zeta}\Psi = -D_{\bar{\theta}}\tilde{\zeta}^{\bar{\theta}}\Psi + \tilde{\zeta}^AD_{A}\Psi.
\end{equation}
Since $D_{\bar{\theta}}^2= D_{\theta}^2=0$ this set of transformations explicitly preserve the constraint (\ref{chiral_superfield}).
We could further rescale the fields and define
\begin{eqnarray}
\Xi &=& \frac{\chi}{\sqrt{e}}, \qquad e(Y+A_0) = \frac{f}{e}, \nonumber \\
\sigma &=& \sqrt{e}\psi,
\end{eqnarray}
so that under reparametrisation,
\begin{eqnarray}
\delta \left(e , \Xi, e(Y+A_0)\right) &=& a\partial_{\tau}\left(e , \Xi, e(Y+A_0)\right) + \dot{a} \left(e , \Xi, e(Y+A_0)\right), \nonumber \\
\delta \sigma &=& a\dot{\sigma}.
\end{eqnarray}
In these new variables, the gravitini transforms as
\begin{equation}
\delta \Xi = 2\dot{\alpha} -i\alpha e^2(Y+A_0) -i\alpha \frac{\bar{\Xi}\Xi}{4e} - it\Xi, \qquad
\delta \bar{\Xi} = 2\dot{\bar{\alpha}} + i\bar{\alpha} e^2(Y+A_0) + i\bar{\alpha} \frac{\bar{\Xi}\Xi}{4e} + it\bar{\Xi},
\end{equation}
where $\alpha = \beta/\sqrt{e} $.
Gauge fixing $\Xi$ and $\bar{\Xi}$ to zero then gives rise to ghost actions
\begin{equation}
\beta(2\partial_{\tau} \pm ie^2(Y+A_0))\gamma.
\end{equation}
Note that automatically we need two sets of ghosts of opposite masses
as is required. Under T-duality along the $Y$ direction where we
exchange $Y+A_0$ for the momentum $k_Y$, these ghosts have opposite
chiralities in the effective theory living in the 1+1 dimensional
intersection domain of the branes. They should be compared with gauge
fixing of the world-sheet local supersymmetries in usual superstring
theories.
These ghosts are the origin of the \emph{bare} CS term and mass term
as is found in the open-string calculation.  We have considered the
special case where $K=1$ where there is only one complex Ramond ground
state fermion. For general $K$ we have $K$ complex fermions. Each of
these fermions possesses local supersymmetries and thus give rise to
$2K$ sets of superghosts.  It is now clear that the extra piece in the
effective potential can be understood entirely in the framework of
effective field theory, and whose coefficient is uniquely fixed, by the coupling of the gauge theory to a $0+1$ supergravity theory. Interestingly, our construction also implies
that the gauge field to which the fermion couples gauges the
R-symmetry of the spontaneously broken local supersymmetries.

\subsection{A remark on half-strings}
We have learnt from the probe calculation that if the mysterious
RR exchange is ignored a D$p$ probe apparently
experiences a nontrivial potential in a
D$(8-p)$ background. The RR exchange generates a balancing force that was
interpreted as the effect of  a \emph{half-string} in \cite{Bergman, Klebanov}.  There are
other arguments that have supported this notion of a
half-string\cite{Banks, Klebanov}. Taking again the example of the
D$p$ probe brane and examining the Chern-Simons terms, there is a term
$\mu_p \int_{\mathcal{M}_{p+1}}  C_{p-1}\wedge F$ (for $p \le 8-p$),  which, upon
integration by parts, becomes
\begin{equation}\label{CS1}
\mu_p \int_{\mathcal{M}_{p+1}} H_{p}\wedge A,
\end{equation}
where $\mathcal{M}_{p+1}$ is the manifold spanned by the probe world-volume .
Since the background D$(8-p)$ brane is magnetically charged with respect
to $H_p$, it can be integrated over the hemisphere $h_p$, transverse to the background brane, upon which the world-volume of the non-compact probe wraps , yielding
$\frac{1}{2}\mu_{8-p}$. The factor of
$\frac{1}{2}$ arises because the integral is over a hemisphere.
Equation (\ref{CS1}) thus becomes
\begin{equation}\label{halfstring1}
\frac{1}{2}\mu_p\mu_{8-p}\int A = \frac{1}{4\pi\alpha'} \int A.
\end{equation}
This is half of the charge carried by the end-point of the
string on the brane induced by the background flux.
Furthermore, as discussed in the previous subsection
the $K$ complex fermionic fields, in $N=8$ quantum
mechanics, gave
rise to a CS term $\frac{K}{2}A_0$ upon being integrated out.
The \emph{bare} term is needed so that the overall coefficient of the CS term is integer valued for consistency.
 This bare term is interpreted as arising from (\ref{CS1})
 where half-strings supply the extra fractional charges.

In perturbative string theory however, there is no notion of fractional string states. These closed string tree-exchange, can be understood as loops in the open string channel, in which usual open string states circulate.  It should be clear from our analysis in the last subsection that 
every term in the low energy effective field theory arises from usual open-string states.


\section{Conclusion}
In this work we have reviewed the problem of string creation and the
remaining questions that have not yet been satisfactorily
answered. Namely, that while a perturbative string calculation yields
a vanishing potential (or a potential due to a stretched string)
between a pair of D$p$ and D$(8-p)$ branes a probe calculation gives a
non-zero potential. The problem persists to M-theory. To yield a
vanishing potential, it was suggested
that there exist half strings
connecting the branes. On the other hand, a careful analysis
in\cite{Divecchia} suggests
a surprising non-zero contribution from the RR sector hitherto not
considered. However, extra duality relations, which are neither
Lorentz invariant nor gauge invariant, were obtained to explain the
unexpected coupling.  These calculations involve a regularisation
procedure to tame the divergences from the ghost sector. To avoid the
use of ghosts which potentially causes confusion, we propose gauge fixing
in a light-cone like gauge. It is a direct closed string version of
Arvis' gauge\cite{Arvis} and is capable of describing these
D$p$-D$(8-p)$ systems. The same conclusion as in \cite{Divecchia}, of
a non-zero RR exchange is reached without encountering
divergences. However, the gauge makes it clear that the extra duality
relations are in fact artifacts of our special gauge choice. Moreover,
in order that the closed string propagator makes sense, the momentum exchanged
along $p_9$ has to vanish in the Euclidean picture, which corresponds
to $p_0$ in the Minkowskian picture. This suggests that the time
direction is special and further Lorentz symmetry is broken before
such exchange is possible. This means that the supergravity solutions
of  parallel D$p$ branes should not be affected by this apparent extra
coupling to a different RR potential.
Finally we discuss how the mismatch in the effective potential between the effective field theory calculation and the corresponding open-string theory calculation can be reconciled. The extra piece in string theory is attributed to
contributions from $\beta \gamma$ ghosts zero modes. The piece is
known as the \emph{bare} term in the field theory, added in by hand
for consistency and supersymmetry. We show that as in string theory,
this term originates from ghosts for gauge fixing spontaneously broken local world-line supersymmetry  of the Ramond ground state fermion. In
any case, the half-tensions arose from quantum effects. The term
\emph{half-strings} is perhaps a misnomer since a classical stringy
object with half of the tension of a fundamental string need never be introduced.


\appendix

\section{Arvis gauge and open string loop}\label{arvis}
A closed string tree exchange can be mapped to a 1-loop open string
diagram by a suitable conformal transformation. More intuitively, the
dynamics of D branes can be described by open strings and the
interaction energy between two D branes can be obtained from the
1-loop vacuum diagram of an open string connecting the two D branes,
using the Coleman-Weinberg formula. To avoid the use of ghosts, we
will introduce the Arvis gauge\cite{Arvis} which is a form of
light-cone gauge where the ``light-cone'' directions $x_0$ and $ x_9$
are allowed to satisfy different boundary conditions.

In the original context, the gauge was used for describing the interaction of two D-particles separated along $x_9$. It is thus convenient to introduce a gauge such that $x_0$ satisfies Neumann boundary condition on both ends of the open string whereas all other coordinates should satisfy Dirichlet boundary condition. This implies that in the conformal gauge the world-sheet fields admit the following Fourier expansion\cite{Arvis}
\begin{eqnarray}
x^i(\sigma, \tau) &=& R^i\sigma + \sqrt{2} \sum_{n\ne 0} \frac{a_n^i}{n} \sin{n\sigma} e^{-in\tau}, \nonumber \\
x^0(\sigma, \tau) &=& q_0 + p^0 \tau+ i \sqrt{2} \sum_{n\ne 0} \frac{a_n^0}{n} \cos{n\sigma} e^{-in\tau}.
\end{eqnarray}

Further gauge conditions
\begin{eqnarray}\label{gauge_cond1}
\dot{x}^9 + x'^0 &=& 0, \nonumber  \\
\dot{x}^0 + x'^9  &=& p^0 + R, \nonumber \\
q^0 = 0,
\end{eqnarray}
are subsequently imposed, which is equivalent to setting
\begin{equation}
a^+_n \equiv a^0_n + a^9_n = 0.
\end{equation}

The Virasoro conditions can then be solved explicitly such that $a^-_n \equiv a^0_n- a^9_n$ can be expressed in terms of the rest of the transverse oscillators.  The Lorentz generators can be built accordingly. Clearly $x^0$ is different from the components $x^i$ due to boundary conditions. For bosonic strings it can be shown that the system enjoys $SO(25)$ rotational invariance for $d=26$.
This can be readily generalised to open superstring theory. The gauge condition (\ref{gauge_cond1}) can be written as
\begin{equation}
\partial_+ x^+ = \partial_-x^- = p^0 + R
\end{equation}
which will make generalisation to closed string pursued in the next section more transparent.

The interaction energy between a D$p$ and a D$(8-p)$ string is then given by
\begin{eqnarray}\label{coleman_weinberg}
&&\int_0^{\infty} \frac{dt}{2t} (8\pi^2\alpha't)^{-1/2} e^{-L^2t/(2\pi\alpha')} \textrm{tr}_{R}(\frac{1+(-1)^{F_0}\Gamma_{\textrm{R}}}{2}e^{-2\pi t(L_0-a_{\textrm{R}})}) \nonumber \\
&&+  \textrm{tr}_{NS}(\frac{1+(-1)^{F_0}\Gamma_{\textrm{NS}}}{2}e^{-2\pi t(L_0-a_{\textrm{NS}})}) \label{trace},
\end{eqnarray}
where $L_0$ is the zero mode of the Virasoro operator, $a_{\textrm{R, NS}}$ are the vacuum contributions to the energies in the respective sector and $\frac{1+(-1)^{F_0}\Gamma_{\textrm{R,NS}}}{2}$ are the GSO projection in the respective sectors. The Gamma matrices $\Gamma_{\textrm{R,NS}}$ are built from the zero modes of the world-sheet fermions. The operator $F_0$ is again the world-sheet fermion number operator. It will be convenient to define $q = \exp(-\pi t)$. When there are exactly 8 relatively transverse directions, all transverse world-sheet fermions in the Ramond sector obtain half-integer moding but integer moding in the NS sector. Therefore tr($(-1)^{F_0}\Gamma_{\textrm{NS}}$) vanishes but  tr($(-1)^{F_0}\Gamma_{\textrm{R}}$) gives a  non-zero contribution.

Adding the contributions together the potential between the branes is given by
\begin{equation}\label{full_sum}
-\int_0^{\infty} \frac{dt}{2t} (8\pi^2\alpha't)^{-1/2}
 e^{-L^2t/(2\pi\alpha')} [\frac{-f_2^8+ f_3^8 \pm f_4^8}{f_4^8}] =
 -\frac{1}{2}T_1L[1\pm 1].
\end{equation}

The third term originates from  tr($(-1)^{F_0}\Gamma_{\textrm{R}}$). The choice of signs corresponds to the possible choices of GSO projections, usually interpreted as the different projections required for branes or anti-branes. However, opposite GSO projections corresponds to a flipped relative orientation of the branes and is equivalent to the two branes crossing each other. Therefore we are led to the same conclusion as in the closed string calculation that a string has to be created whenever the branes cross each other.

\section{Open string partition sum in conformal gauge and regularisation}\label{conformal}

A crucial part in the calculation of the interaction potential between
the branes is to evaluate the partition sum
\begin{equation}\label{partition}
\textrm{tr}(\frac{(1+(-1)^F)}{2}q^{2N}).
\end{equation}
The sum is straight forward in Arvis gauge as is presented in the previous section.  In conformal gauge, the residual symmetry requires the introduction of ghosts. The partition sum should therefore also include their contribution. The residual Virasoro symmetry calls for a pair of anti-commuting $bc$ ghosts and the world-sheet supersymmetry calls for a pair of commuting $\beta \gamma$ ghosts.
These ghosts can be expanded in Fourier modes and be quantised. Their modes satisfy the following (anti-)commutation relations\cite{GSW}
\begin{eqnarray}
b_{++} = \sum_{n} b_{ n} \exp{-in\sigma_{+}}, &\qquad& c^{+} =
 \sum_{n} c_{ n} \exp{-in\sigma_{+}}, \nonumber \\
 \beta_{3/2} = \frac{1}{\sqrt2} \sum_{n}  \beta_{\nu + n}
 \exp{-i(n+\nu)\sigma_{+}}, &\qquad&  \beta_{-1/2} = \frac{1}{\sqrt2}
 \sum_{n}  \gamma_{\nu + n} \exp{-i(n+\nu)\sigma_{+}} \nonumber \\
 \{c_s, b_t\} = \delta_{s+t}, &\qquad&    \{c_s, c_t\}= \{b_s,
 b_t\}=0,  \nonumber \\
\lbrack\gamma_s,\beta_t\rbrack = \delta_{s+t}, &\qquad&
 \lbrack\gamma_s, \gamma_t\rbrack =
 \lbrack\beta_s,\beta_t\rbrack = 0,
\end{eqnarray}
where $n$ are integers and $\nu = 0$ in the R sector and $1/2$ in the NS sector.
The ground states are defined such that they are annihilated by the positive modes of the ghosts.
The zero modes of the $bc$ ghosts, and the $\beta\gamma$ ghosts in the R sector, generate degenerate ground states. In particular, since the $\beta\gamma$ commutes, their zero modes generate an infinite tower of states. From the (anti-)commutation relations we can treat $c_0$ and $\gamma_0$ as a creation operators, and $b$ and $\beta_0$ as annihilation operator. Therefore the degenerate ground states can take the form
\begin{equation}
\gamma_0^n c_0^l |0,k_\mu>,
\end{equation}
where $ b_0|0,k_\mu> = \beta_0 |0,k_\mu>=0$. Also, $n \ge 0$ and $l\in \{0,1\}$. The ground state is infinitely degenerate. Physical states have to satisfy
\begin{equation}
b_0|\textrm{phys}> = \beta_0 |\textrm{phys}>=0.
\end{equation}
In addition, GSO projection should remove half of the ghost states. A physical state should thus satisfy also
\begin{equation}
(1-(-1)^{F+G_{\beta\gamma}}) |\textrm{phys}> = 0,
\end{equation}
where $F$ is the world-sheet fermion number and $G_{\beta\gamma}$ the $\beta\gamma$ ghost number. Explicitly, in the R sector
\begin{equation}
G_{\beta\gamma} = \gamma_0\beta_0 + \sum_{n>0}\gamma_{-n}\beta_{n} + \beta_{-n}\gamma_{n},
\end{equation}
and similarly in the NS sector where the zero modes are absent. We are
then ready to compute the partition sum. Putting in the GSO projector
explicitly as in (\ref{partition}) the sum breaks down into four
different terms, two in the NS sector and another two in the R sector.
The trace over the world-sheet fermions implies anti-periodic boundary
conditions in the world-sheet time direction. The insertion of $(-1)^F$
in the trace corresponds to flipping to periodic boundary condition.
Therefore we can denote the contribution of the four terms according
to the boundary conditions along $(\sigma,\
\tau)$ of the world-sheet supersymmetry current $T_F (\sigma_+)= \psi
. \partial_{+}X$ as (NS,NS), (NS,R), (R,NS) and (R,R).  It is
important that the boundary conditions of the ghosts should match
those of the corresponding world-sheet current. i.e. The $bc$ ghosts
should have the same boundary conditions as the energy momentum tensor
$T$ and the $\beta\gamma$ ghosts take the boundary conditions of the
world-sheet supersymmetry current. Therefore the $bc$ ghosts always
have periodic boundary conditions along both $\sigma$ and
$\tau$. Given that they anti-commute, a factor of $(-1)^{G_{bc}}$ is
inserted in all the traces, where $G_{bc}$ is the $bc$ ghosts number.
For the $\beta\gamma$ ghosts their boundary conditions are as
specified in the four sectors. They are commuting and so to obtain
anti-periodic boundary conditions along $\tau$ we have to insert
$(-1)^{G_{\beta\gamma}}$ in the trace.  The partition sum in the
(NS,NS) and (NS,R) sectors immediately reduce to those obtained in the
gauge fixed calculation, where only the non-light-cone directions
contribute. In the (R,NS) and (R,R) sector however the situation is
rendered more complicated by the infinite tower of degenerate ground
states due to the presence of $\beta\gamma$ zero modes. Since the
trace over all states factorises for each pair of creation and
annihilation operator, we can do the sum separately for the
$\beta\gamma$ ghosts. In the (R,NS) sector
\begin{equation}
\textrm{tr}((-1)^{G_{\beta\gamma}}q^{2N_{\beta\gamma}}) = \prod^{\infty}_{n=0} (\sum_m (-1)^m q^{2nm}) .
\end{equation}
The factor at $n=0$ gives an infinite alternate series $1-1+1....$. To regularise we take the trace
\begin{equation}\label{regularise1}
\textrm{tr}((-x)^{G_{\beta\gamma}}q^{2N_{\beta\gamma}}),
\end{equation}
and only take the limit $x \to +1$ at the end. The infinite series then reads
\begin{equation}\label{regularise2}
\sum_{n=0}^\infty (-x)^n = \frac{1}{1+x}.
\end{equation}
This gives a factor of $1/2$ in the limit $x \to +1$.  Similarly in
the (R,R) sector the trace should be
\begin{equation}
\textrm{tr}(q^{2N_{\beta\gamma}})=
\prod_{n=0}^{\infty} \sum_m {q^{2nm}}.
\end{equation}
Clearly the zero modes contribution  diverges. Applying the same
regularisation as in (\ref{regularise1}) by tracing over
$\textrm{tr}((x)^{G_{\beta\gamma}}q^{2N_{\beta\gamma}})$, the zero
modes give $1/(1-x)$ and diverges in the limit $x\to +1$.

Consider then the trace over the world-sheet fermions. Their zero modes satisfy the Clifford algebra. We can define (for simplicity we consider a Euclidean space)
\begin{equation}
\Psi_0^{\pm a} = \frac{\psi^{2a}_0 \pm i \psi^{2a+1}_0}{2},
\end{equation}
where $a={0,1,... n/2-1}$,  when there are $n$ Neumann-Neumann or Dirichlet-Dirichlet directions. Since they satisfy anti-commutation relations
\begin{equation}
\{\Psi^{a+}_0, \Psi^{b-}_0\} = \delta_{ab},
\end{equation}
they can be treated as creation and annihilation operators after defining a ground state annihilated by all $\Psi^{a-}_0$.
Consider the open strings connecting a D0 and a D8. In such cases
$n=2$. The $\Psi_0^{\pm}$ generate two degenerate ground states $|\downarrow>$
and $\Psi_0^+|\downarrow>= |\uparrow>$.
In the (R,NS) trace the two world-sheet fermionic ground states give a factor of $q^0+ q^0=2$. In the (R,R) sector the trace is
\begin{equation}
\textrm{tr}((-1)^Fq^{2N}).
\end{equation}
The zero modes part of $(-1)^F$ can be written as $(-1)^{\Psi^+_0\Psi^-_0}$.
The zero modes give $<\downarrow|\downarrow>-<\uparrow|\uparrow>=1-1=0$. To \emph{regularise} the sum as for the ghosts we can instead take the trace
\begin{equation}
\textrm{tr}((-x)^Fq^{2N}),
\end{equation}
after which we should take the limit $x \to +1$. The zero modes
contribution is then equal to $1-x$. The total contribution of $\beta\gamma$
ghosts and world-sheet fermion zero modes are thus given by
\begin{equation}
\lim_{x\to +1}\frac{1-x} {1-x} = 1.
\end{equation}


\section{A note on monodromy in D(-1)- D7 system}
The D(-1)- D7 system\footnote{We thank
    J. Maldacena for suggesting this related problem.} is
    another example where there are eight relatively transverse
    dimensions between the branes and we should expect the counterpart
    of string creation to occur here. The configuration has
    codimension two, compared to D$p$-D$(8-p)$ systems which have
    codimension one. Instead of crossing the two branes there is a two
    plane on which we can move the D-instanton relative to the
    D7. Therefore the natural counterpart of string creation in this
    configuration should be a monodromy in the complexified dilaton
    $\tau = c_0 + i2\pi e^{-\phi}$ when the D-instanton encircles the
    D7 brane. To exhibit this effect, we evaluate
\begin{equation}
<D-1| \Delta |D7>.
\end{equation}
This is analogous to what has been done for other D$p$- D$(8-p)$ systems. The result is
\begin{equation}
\frac{A}{2\pi} \int_0^{\infty}\frac{dt}{t} e^{-|z|^2/(2\alpha't)} (1\pm 1),
\end{equation}
where $z$ is the complex coordinate of the two totally transverse dimensions. The integral should yield a harmonic function in $z$. Now, given that under a parity transformation that transforms $z \to e^{i\pi}z$, the result jumps from zero to one. The integral is thus
\begin{equation}
<D-1| \Delta |D7> = \frac{A}{\pi} \log{z}.
\end{equation}

\acknowledgments
I would like to thank M. Green for his guidance and for
the useful discussions. I would also like to thank J. Maldacena,
D. Tong and L. I. Uruchurtu for comments and suggestions.
I am also indebted to the Gates Cambridge Trust and ORSAS UK for
financial support.

\end{document}